\documentclass{ws-ijmpa}

\begin{document}

%

\def\nocropmarks{\vskip5pt\phantom{cropmarks}}


%

\markboth{Y.-Y. Keum}
{Angular distribution analysis of $B \to J/\psi \, K^{*}$ and 
resolving discrete ambiguities on
the determination $\phi_1$.}

%
\catchline{}{}{}
%

\setcounter{page}{1}

\title{Angular distribution analysis of $B \to J/\psi \, K^{*}$ and 
resolving discrete ambiguities in
the determination of $\phi_1$
}

\author{\footnotesize  
H.-Y. Cheng and Y.-Y. Keum\footnote{Presented by Y.-Y. Keum at the
3rd Circum-Pan-Pacific Symposium on High Energy Spin Physics
(Bejing, China, Oct.8-13, 2001)}}

\address{Institute of Physics, Academia Sinica, Address\\
Nankang Tapei, Taiwan 115,Republic of China
}

\author{K.-C. Yang}

\address{Department of Physics, Chung Yuan Christian University\\
Chung-Li, Taiwan 320, Republic of China
}

\maketitle


\begin{abstract}
We discuss the angular distribution analysis of 
$B\to J/\psi \, K^{*}$ decays
and a way to resolve discrete ambiguites 
in the determination of the unitary
triangle $\phi_1(=\beta)$. We study the status of factorization
hypothesis in the color-suppressed B meson decays: $B\to J/\psi
K^{(*)}$ within
the general factorization approach 
and QCD-factorization method.

\end{abstract}

\section{Introduction}	

Among a hundred nonleptonic two body decays of B mesons,
the process $B \to J/\psi K^{*}$ has lots of interests in
many aspects: First of all, it was firstly observed color-suppressed 
process with a large branching ratio in B meson decays.$^{1-5}$
 The vector-vector decays 
$B^0 \to J/\psi K^{*0}(K^{*0} \to K_s^0\pi^0)$
is a mixture of CP-even and CP-odd eigenstates since it can proceed
via an S,P,D wave decays. By using both angular and time distribution
analysis, we can separate the CP-even one from CP-odd eigenstate and
determine the angle $\phi_1(=\beta)$ of the unitarity triangle
without any dilution effects in a manner similar to 
which the CP-odd eigenstate $B^0 \to J/\psi K^{0}$ is used.
In addition, the angular distribution analyses on 
both $B^0\to J/\psi K^{*}$ and 
$B_s^0 \to J/\psi \phi$ can be used to resolve 4-fold ambiguities
in the measurement of $sin 2\phi_1$.

The recent measurement by BaBar$^{4}$ 
has confirmed the earlier CDF$^{2}$ observation that there is a
nontrivial strong phase difference between polarized amplitudes
indicating final-state interactions. However no such evidence
has been seen yet by CLEO$^{3}$ and Belle$^{4}$.
It is interesting to check if the current QCD-approaches for B
hadronic decays predicts a departure from factorization. Therefore,
the measurements of various helicity amplitudes in $B \to J/\psi
K^{*}$ decays will provide a powerful tool for testing factorization 
and differentiating various theorical models$^{6-9}$ in which 
the calculated nonfactorizable term have real and imaginary parts .

On the other hand, precise measurement of the CP asymmetry in the 
$B \to J/\psi K^{(*)}$ decays is important for new physics search 
by comparing with one of $B \to \phi K$ with high degree of
accuracy.$^{10,11}$ 
This measurement is experimentally accessible 
at the early stage of the asymmetric $B$ factories. The $B\to \phi K$
decays arise from penguin (loop) effects, 
while the $B\to J/\psi K^{(*)}$ decays
involve dominant tree amplitudes. 
The search for different CP asymmetries in
the $B \to J/\psi K^{(*)}$ and $\phi K$ decays, with the common source from
$B^0$-$\bar{B}^0$ mixing, provides a promising way to discover
new physics $^{12,13}$: a difference of
$|A_{CP}(J/\psi K^{(*)}) - A_{CP}(\phi K)| > 5 \%$ would be an indication
of new physics.

\section{Angular Anaysis and Resolving Discrete Ambiguities}
The measurement of $sin2\beta$ has a four-fold ambiguity:
${\phi_1, \pi/2-\phi_1, \pi+\phi_1,3\pi/2 -\phi_1}$
with $0<\beta<2\pi$.
In order to resolve this ambiguity, one need to determine the signs
of $cos2\phi_1$ and $sin\phi_1$ in addition to the value of $sin2\phi_1$.
\subsection{Determination of sign($cos2\phi_1$)}
Using interference between opposite CP amplitudes in $J/\psi K^{*}$
and $J/\psi \phi$ can help to determine the sign of $cos2\phi_1$.
The interference term between CP-even and CP-odd amplitude which
can be obtained from the transversity analysis contains a term
in $cos2\phi_1$. For instance,
\begin{eqnarray}
Im[A_{\perp}(t)A_{||}^{*}(t)] \sim && 
Im[A_{\perp}(0)A_{||}^{*}(0)] \,\, cos\triangle mt \nonumber \\
&& - Re[A_{\perp}(0)A_{||}^{*}(0)] \eta \, cos2\phi_1 \,\,
sin\triangle mt. \label{eq1}
\end{eqnarray}
Observables in transversity frame for $J/\psi(K^{(*0)})_{CP}$ is given
by in Table I.

\begin{table}[htbp]
\ttbl{30pc}{} 
{\begin{tabular}{ccc}\\ \hline
Time-dependent Obs. & Time-dependence & Time-independence \\ \hline
$|A_{||}|^2$ (CP=$+$) &   &   \\ 
$|A_{\perp}|^2$ (CP=$-$) & $sin\triangle mt$ & $sin\,2\phi_1$  \\
$|A_{0}|^2$ (CP=$+$) &   &   \\ \hline
$Re[A_{||}A^{*}_0]$ & constant & $cos[\phi(A_{||})-\phi(A_0)]$ \\
        & $sin \triangle mt$ & $cos[\phi(A_{||})-\phi(A_0)]\,\, sin\,2\phi_1$ \\
\hline
$Im[A_{\perp}A^{*}_{||}]$ & $sin\triangle mt$ & 
$cos[\phi(A_{\perp})-\phi(A_{||})]$ $cos\,2\phi_1$ \\
        & $cos \triangle mt$ & $sin[\phi(A_{\perp})-\phi(A_{||})]$ \\
\hline
$Im[A_{\perp}A^{*}_{0}]$ & $sin\triangle mt$ & 
$cos[\phi(A_{\perp})-\phi(A_{0})]$ $cos\,2\phi_1$ \\
        & $cos \triangle mt$ & $sin[\phi(A_{\perp})-\phi(A_{0})]$ \\
\hline
\end{tabular}}
\end{table}
Since experiments measured
interference terms in the angular distribution with
Re$(A_\|A_0^*)$, Im$(A_\perp A_0^*)$ and Im$(A_\perp A^*_\|)$,
there exists a phase ambiguity with $\phi(A_0)=0$:
 \begin{eqnarray}
 \phi_\| & \to& -\phi_\|, \nonumber \\
 \phi_\perp & \to& \pm\pi-\phi_\perp,  \\
 \phi_\perp-\phi_\| & \to& \pm \pi-(\phi_\perp-\phi_\|). \nonumber
 \end{eqnarray}
It is easy to check that there is a sign ambiguity on 
$cos[\phi(A_{\perp}) - \phi(A_{||})]$ and on $cos\phi(A_{\perp})$.
Therefore a sign ambiguity on $cos\,2\phi_1$ remains.

There are two solutions for the relative phases according to 
BaBar measurement$^{4}$ as an example  :
 \begin{equation}
 \phi_\perp=-0.17\pm0.17\,, \qquad\qquad \phi_\|=2.50\pm 0.22\,,
 \qquad \Rightarrow~~ |H_+|<|H_-|,
 \label{I}
 \end{equation}
where the phases are measured in radians. The other allowed
solution is
 \begin{equation}
 \phi_\perp=-2.97\pm0.17\,, \qquad\qquad \phi_\|=-2.50\pm
 0.22\,,\qquad \Rightarrow~~ |H_+|>|H_-|.
 \label{II}
 \end{equation}
As pointed out by Suzuki $^{14}$, the solution (\ref{I}) indicates
that $A_\|$ has a sign opposite to that of $A_\perp$ and hence
$|H_+|<|H_-|$, in contradiction to what expected from
factorization. Therefore, we prefer to solution (\ref{II}) to compare with
the factorization approach. Obviously there is a 3-$\sigma$ effect
that $\phi_\|$ is different from $\pi$ and this agrees with the
CDF measurement$^{3}$. However, such an effect is not observed by Belle$^{5}$
and CLEO$^{2}$ (see Table 4). 

In fact, We can determine unambiguously the strong phase of 
$cos\,2\phi_1$ term
by studying the angular distribution analysis of $B_s \to J/\Psi \phi$
with  SU(3) flavour symmetry $^{15}$.

\subsection{Determination of $sign(sin\phi_1)$}

The determination of $sign(sin\phi_1)$ leaves the ambiguity of
$\phi_1 \to \pi+\phi_1$. However, it needs some model-dependent input.
By comparing the coefficients of $sin \triangle mt$ of $J/\Psi K_s^0$
vs $D^{+}D^{-}$, we obtain :
\begin{eqnarray}
S_{J/\Psi K_S^0} &=& - sin\,2\phi_1, \nonumber \\
\cr
S_{D^{+}D^{-}} &=& 
\left[ {sin\,2\phi_1 -2 |R_{DD}| sin\phi_1 \,\, cos\delta_{DD} } \over
{1 + |R_{DD}|^2 - 2 |R_{DD}| cos\phi_1 cos\delta_{DD} }  \right]
\end{eqnarray} 
which can give the $sign(sin\phi_1)$
if $sign(cos\,2\phi_1)$ and $sign(cos\delta_{DD})$ are known.
The $sign(cos2\phi_1)$ could be determined by method(2-1).

The determination of $sign(cos\delta_{DD})$ need model-dependent
input.
\begin{equation}
S_{J/\Psi \, K_S^0} + S_{D^{+}D^{-}} = 2 \,\, |R_{DD}| \,\, cos\delta_{DD}
cos\,2\phi_1 \,\, sin\phi_1.
\end{equation}
If $cos\delta_{DD} > 0$, we obtain the relation :
\begin{equation}
sign[S_{J/\psi\, K_S^0} + S_{D^{+}D^{-}}] = sign[cos\,2\phi_1 \,\, sin\phi_1]
\end{equation}
In Standard Model $sign(cos\,2\phi_1)$ is positive.

\section{Test of Factorization in $B \to J/\psi K^{*}$ decay}
By using the angular distribution analysis in the transversity basis,
we can measure precisely both their magitudes and phases of
the three different helicity amplitudes, denoted by $H_0, H_{-}$, and
$H_{+}$. These observations can provide a crucial way to test 
not only the naive factorization method but also 
recent improved QCD-approaches in which non-factorizable term can be
calculable. Also it would answer the question of the existence of 
the final state interactions, which is strong enough to flip the quark
spin in color-suppressed B decays$^{14}$. 
 
\subsection{General Factorization Approach for $J/\psi K^{*}$$^{16-19}$}
It has been well known that the factorization approach (naive or
generalized) fails to explain simultanuously 
the production ratio $R={\cal
 B}(B\to J/\psi K^*)/{\cal B}(B\to J/\psi K)$ 
and the fraction of longitudinal polarization $\Gamma_L/\Gamma$ 
in $B\to J/\psi K^*$ decay$^{16}$.

\begin{table}[htbp]
\ttbl{30pc}{}
\caption{The ratio of vector meson to pseudoscalar production $R$
and the longitudinal polarization fraction $\Gamma_L/\Gamma$ in
$B\to J/\psi K^{(*)}$ decays calculated in two representative
form-factor models using the factorization hypothesis.
 \label{tab2}}
\begin{center}
{\begin{tabular}{lcccccc} \hline
& & & \multicolumn{4}{c}{Experiments}  \\ \cline{4-7}
 & \raisebox{1.5ex}[0cm][0cm]{BB$^{20}$} & \raisebox{1.5ex}[0cm][0cm]{MS$^{21}$}
& CDF & CLEO & BaBar  & Belle \\
\hline
  $R$  & 3.40 & 3.11 & $1.53\pm 0.32$ & $1.45\pm 0.26$ 
& $1.38\pm 0.11$ & $1.43\pm0.13$ \\
$\Gamma_L/\Gamma$& 0.47 & 0.46 & $0.61\pm 0.14$ & $0.52\pm 0.08$
 & $0.60\pm 0.04$ & $0.60\pm0.05$ \\ \hline
\end{tabular}} 
\end{center}
\end{table}
In the general factorization approach, non-factorizable term is directly
propotional to the factorizable piece and have the same phase as
factorized one.
So the theoretical difficulty can be understandable 
because we assumed the parameter $a_2$ to be universal 
according to the factorization hypothesis,
namely $a_2^{h}(J/\psi K^*)=a_2(J/\psi K)$
where $h=0,+,-$  refer to the helicity states $00$, $++$ and $--$
respectively.
In this case, the amplitudes are relatively real and there is no
significant signature of the final state interaction, which is agreed
with experimantal results of CLEO and Belle, but contradicted to BaBar
and CDF results. 

\subsection{QCD-improved factorization approach for $J/\psi K^{*}$ $^{22,23}$}
The QCD-improved factorization approach$^{8}$ 
allows us to compute the nonfactorizable corrections
in the heavy quark limit since only hard interactions between the
$(BV_1)$ system and $V_2$ survive in the $m_b\to\infty$ limit.
In this approach, the light-cone distribution amplitudes (LCDAs) 
play an essential role. It is shown that non-factorizable
terms contribute differently to each helicity amplitude and to
different decay modes so that $a_2^0(J/\psi K^*)> 
a_2^+(J/\psi K^*)\neq a_2^-(J/\psi K^*)$ 
and $a_2(J/\psi K)> a_2^h(J/\psi K^*)$. 
With non-relativistic and asymptotic type $J/\psi$ LACDs, and upto
twist-3 LCDAs for $K^{*}$,
we find that (i) for $B \to J/\psi K$, twist-3 hard spectator
interaction are equally important as twist-2 contributions,
(ii) however, for $B \to J/\psi K^{*}$, the spectator and final state
interactions from leading twist contributions play an important role
in the dominant lognitudinal component, 
which is safe from the infrared divergence and induce
$|a^0(J/\psi K^{*})| \sim 0.14$ different from  $|a^0(J/\psi K)| \sim 0.2$.

\begin{table}[htbp]
\ttbl{30pc}{} 
\caption{Normalized spin amplitudes and their phases (in radians)
in $B\to J/\psi K^*$ decays calculated in various form-factor
models using QCD factorization. The branching ratios given in the
Table are for $B^+\to J/\psi K^{*+}$. For comparison, experimental
results form CDF, CLEO, BaBar and Belle are also exhibited.
\label{tab3}}
\begin{center}
{\begin{tabular}{l c c c c c c} \hline
Model & $|\hat A_0|^2$ &  $|\hat A_\bot|^2$ &  $|\hat A_\||^2$ &
$\phi_\bot$ & $\phi_\|$ & ${\cal B}(10^{-3})$ \\ \hline
 BSWI$^{6}$ & 0.43 & 0.33 & 0.24 & $-3.05$ & $-2.89$  & 0.76 \\
 BSWII$^{25}$ & 0.38 & 0.36 & 0.26 & 3.13 & $-3.12$ & 0.73 \\
 LF$^{26}$ & 0.41 & 0.34 & 0.25 & $-3.09$ & $-2.95$ & 0.69 \\
 NS$^{27}$ & 0.40 & 0.34 & 0.25 & $-3.10$ & $-2.99$ & 0.70 \\
 Yang$^{28}$ & 0.38 & 0.36 & 0.25 & $-3.12$ & $-3.11$ & 0.64 \\
 BB$^{20}$ & 0.41 & 0.34 & 0.25 & $-3.04$ & $-3.05$ & 0.77 \\
 MS$^{21}$ & 0.40 & 0.35 & 0.25 & $-3.08$ & $-3.05$ & 0.75 \\
YYK$^{17}$ & 0.44 & 0.32 & 0.23 & $-2.99$ & $-2.95$ & 0.84  \\
 \hline
 CLEO & $0.52\pm 0.08$ & $0.16\pm 0.09$ & & $-3.03\pm 0.46$ &
 $-3.00\pm 0.37$ & $1.41\pm 0.31$ \\
 CDF & $0.59\pm 0.06$ & $0.13^{+0.13}_{-0.11}$ & $0.28\pm0.12$
 & $-2.58\pm 0.54$ & $-2.20\pm0.47$  &   \\
 BaBar & $0.60\pm 0.04$ & $0.16\pm 0.03$ & $0.24\pm 0.04$ &
 $-2.97\pm 0.17$ & $-2.50\pm 0.22$ & $1.37\pm0.14$ \\
 Belle  & $0.60\pm0.05$ & $0.19\pm0.06$ & & $-3.15\pm0.21$ &
 $-2.86\pm0.25$ & $1.29\pm0.14$ \\ \hline
\end{tabular}}
\end{center}
\end{table}
As shown in Table 4, we obtained 
small $a_2^{h}$ which can explain only half of the
data for the branching ratio. 
We also got relatively small fraction of the longitudinal 
polaization component, but large fraction for $|\hat{A}_{\perp}|^2$.

From somehow negative results, we  conclude that it is needed 
to understand more correctly the LCDAs of heavy ($c\bar{c}$)-state 
and the power $\Lambda/m_b$ corrections within QCD-factorization
method.

To get more understanding on factorization in color-suppressed decays
including charmonium states,
we suggest that the study on $B \to \eta_c K^{(*)}$ will provide
a good test of the factorization hypothesis.$^{24}$
We expect $Br(B\to K^+\eta_c)=(1.14\pm0.31)\times 10^{-3}$, which can be
observed in near future.
 
\section*{Acknowledgements}
It is a great pleasure to thank B.-Q. Ma for the invitation to 
this wonderful symposium and for his hospitality.
I wish to acknowledge  joyful discussions with H.-n. Li, X.-Y. Pham, 
and A.I. Sanda.
This work was supported in part by the National Science Council of
R.O.C. under Grant Nos. NSC90-2112-M-001-047, 
NSC90-2112-M-033-004 and NSC90-2811-M-001-019.

\end{document}